\documentclass[journal]{IEEEtran}
\usepackage{amsmath,lipsum}
\usepackage{cuted}
\ifCLASSINFOpdf
\else
\usepackage[dvips]{graphicx}
\fi
\usepackage{url}

\usepackage{graphicx}
\usepackage{cite}
\usepackage{picinpar}
\usepackage{amsmath}
\usepackage{amsthm}
\usepackage{url}
\usepackage{flushend}
\usepackage[latin1]{inputenc}
\usepackage{colortbl}
\usepackage{soul}
\usepackage{multirow}
\usepackage{pifont}
\usepackage{color}
\usepackage{alltt}
\usepackage[hidelinks]{hyperref}
\usepackage{enumerate}
\usepackage{siunitx}
\usepackage{breakurl}
\usepackage{epstopdf}
\usepackage{pbox}
\usepackage{algorithm} 
\usepackage{algorithmic} 
\usepackage{amssymb}
\usepackage{subfigure}
\usepackage{booktabs} 
\usepackage{balance}

\newtheorem{prop}{Proposition}
\newtheorem{theorem}{Theorem}

\definecolor{rev}{rgb}{1.00,0.00,0.00}    

\begin{document}
	\title{High-Order Sinusoidal Modeling for Precise Parameter Refinement}
	\author{ Shaowen Chen and Tomoki Toda
		\IEEEcompsocitemizethanks{\IEEEcompsocthanksitem This work was partly supported by JST AIP Acceleration Research
			JPMJCR25U5, Japan.
			\IEEEcompsocthanksitem Shaowen Chen is with the Graduate School of Informatics, Nagoya University, Nagoya 464-8601, Japan (e-mail:shaowen.chen@g.sp.m.is.nagoya-u.ac.jp)
			\IEEEcompsocthanksitem Tomoki Toda is with the Information Technology Center, Nagoya University,
			Nagoya 464-8601, Japan (e-mail:tomoki@icts.nagoya-u.ac.jp)
	}}
	\maketitle
	
	\begin{abstract}
		
		Although the amplitude estimation of sinusoids is a linear and tractable problem when their frequencies are known, the accuracy of signal modeling is fundamentally limited by the inevitable frequency mismatches produced by existing methods. Quasi-harmonic modeling (QHM) and its extensions (aQHM and eaQHM) offer the ability to refine frequency estimates while extracting amplitudes; however, their extraction and correction capabilities remain limited. In this paper, we first revisit QHM methods from the perspective of frequency mismatches, and reveal that their performances are essentially constrained by the order of the model. Specifically, QHM methods are linear models and thus incapable of capturing higher-order frequency variations, leading to suboptimal modeling accuracy. Second, to address this limitation, we propose a higher-order modeling framework that enables more precise approximation of signal components. Eventually, we qualitatively characterize the optimization problem and employ the Levenberg--Marquardt (LM) algorithm for parameter estimation. As a result, the proposed method allows for significantly improved estimation of amplitudes, frequencies, and phases of sinusoidal components, enabling more accurate signal modeling. Experiments on both simulation study and speech signals demonstrate the superiority of the high-order model in frequency refinement and parameter extraction and its robustness under additive noise conditions.
	
	\end{abstract}
	
	
	\markboth{IEEE TRANSACTIONS ON SIGNAL PROCESSING}
	{}
	
	\definecolor{limegreen}{rgb}{0.2, 0.8, 0.2}
	\definecolor{forestgreen}{rgb}{0.13, 0.55, 0.13}
	\definecolor{greenhtml}{rgb}{0.0, 0.5, 0.0}
	
	\section{Introduction}
	\label{S:1}
	\IEEEPARstart{S}{inusoids} are ubiquitous in the world and usually constitute fundamental components of various real-world signals \cite{2010Flandrin}, such as audio (music, speech) \cite{Qua1986Speech}, radar \cite{Ameri2019radar}, and vibration signals \cite{yang2014general}. In practice, these signals are inevitably perturbed by noise or inherently contain stochastic components. Thus, the observed signals $x(t)$ can be modeled as 
	\begin{align} 
	x(t)= \sum\limits_{k=1}^{K}{x_k(t) + \varepsilon (t)} 
	= \sum\limits_{k=1}^{K}{A_k(t) e^{i \varphi_k(t)} + \varepsilon (t)}
	\label{eq1}
	\end{align}
	where $K$ is the number of fundamental components, $A_k(t)$, $\varphi_k(t)$, and $\varphi'_k(t)$ denote the instantaneous amplitude, instantaneous phase, and instantaneous frequency of the $k$-th component $x_k(t)$, respectively, and $\varepsilon (t)$ is noise. In various fields, such as audio and speech processing \cite{rao2000decomposing},\cite{pantazis2010adaptive} and vibration monitoring \cite{2016Wang}, \cite{2018Wang}, significant emphasis is placed on signal modeling, including analysis and synthesis. During analysis, parameters, such as amplitude, frequency, and phase, are extracted to reveal signal structure, while synthesis uses them to reconstruct or decompose signals. Obviously, the quality of synthesis depends heavily on the accuracy of these parameters, making precise extraction a key research focus.

	Initially, spectrogram-based methods, such as the short-time Fourier transform (STFT) \cite{oberlin2014fourier}, Wigner-Ville distribution \cite{boashash1987efficient}, and wavelet transform, were introduced to visualize the spectral structure of signals by converting them from the time domain to the time--frequency (TF) domain. Based on the TF representations of these methods, signal ridges can be detected for extracting the corresponding sinusoidal parameters of each component\textbf{}. However, they are often limited by the uncertainty principle, which restricts simultaneous high-resolution localization in both time and frequency, indispensably impairing the accuracy of ridge detection and deteriorating the parameter extraction accuracy. Therefore, various postprocessing techniques, including the reassignment method (RM) \cite{1995Auger},\cite{2013Auger} and synchrosqueezing transform (SST) \cite{2011Daubechies},\cite{2013Thakur}, were proposed to sharpen the TF representations for more accurate ridge detections. Additionally, synchrosqueezing has been extended to STFT \cite{2014Herrera} and S-transform \cite{2016Huang} frameworks, and is widely applied in various fields \cite{2014Wu},\cite{2013Wu},\cite{2016Wu}. Nevertheless, accurate parameter estimation still remains difficult due to the inability of these methods to demodulate the signal. 
	
	To further improve the TF resolution, the second-order synchrosqueezing transform (2nd-SST) was proposed to enhance instantaneous frequency (IF) estimates to increase the TF resolution \cite{2015Oberlin}. Subsequently, the 2nd-SST was further extended to high-order versions, enabling a more sharpened TF representation of modulated signals \cite{2017Pham}. Besides, synchroextracting transform \cite{2017Yu} and multi-SST \cite{2018Yu} were proposed with higher TF resolution. Additionally, statistic SST (Stat-SST) was also proposed to achieve a significantly high TF resolution by identifying the IF bands and aligning the IF estimates within IF bands to their fixed points, while also maintaining good reconstruction capability \cite{chen2023instantaneous}. However, its frequency resolution is ultimately limited by the axis resolution, i.e., $\frac{f_\text{s}}{N_{\text{FFT}}}$, where $f_\text{s}$ and $N_{\text{FFT}}$ are the sampling rate and the number of points used in Fourier transform. Obviously, setting $N_{\text{FFT}} = f_\text{s}$ for high-sampling-rate signals, such as speech signals, is impractical, as it incurs an extremely large computational cost.

	In addition to spectrogram-based methods mentioned above, increasing attention has been devoted to parametric methods, with a particular focus on improving parameter extraction accuracy. A representative approach is the sinusoidal model, which models signals as sparse parametric sinusoids by tracking spectral ridges, i.e., framewise amplitude, frequency, and phase. These detected parameters are then sophisticatedly interpolated to their instantaneous forms to capture intra-frame variations, improving reconstruction accuracy. However, the estimation remains limited by spectrogram resolution and the assumption of intra-frame stationarity. To mitigate this, \cite{Stylianou1998on} investigated various amplitude models and showed that time-varying amplitude models significantly reduce estimation error. Subsequently, the quasi-harmonic model (QHM) \cite{pantazis2008properties} applied a complex time-varying amplitude model to speech modeling, improving both amplitude estimation accuracy and frequency correction through least-square (LS) optimization, though the frequency remained constant within frames. Iterative QHM \cite{pantazis2010iterative} was proposed to refine the frequency estimates through iterations; nevertheless, the iterations still relied on the intra-frame stationarity, limiting the correction accuracy. Accordingly, the adaptive QHM (aQHM) adopted a nonlinear phase function to better handle frequency variations. Furthermore, to enhance the adaptability to rapidly varying amplitudes from signals, such as singing voice, an adaptive amplifier was introduced to develop an extension of aQHM (eaQHM) \cite{kafentzis2012extension}, improving the accuracy of the extraction of amplitude, phase, and frequency.

	Despite the significant improvements in parameter estimation achieved by the QHM family, enabling accurate reconstruction of signal components, these methods still face notable limitations. For instance, although iterative schemes can gradually reduce frequency estimation errors, their performance substantially degrades when large initial frequency estimate errors exist, which is often inevitable in practice, particularly for harmonic signals where closely spaced components may interfere with each other. In addition, QHM is sensitive to noise, which is also frequently unavoidable. Under noisy conditions, the LS estimation of complex amplitudes becomes less reliable, further impairing the accuracy of frequency estimation. Therefore, the development of a robust method capable of accurately estimating the amplitude, frequency, and phase of signal components under noisy conditions and frequency deviations is of great significance.
	
	In this paper, the theory of QHM framework is revisited in the time domain, and its frequency correction mechanism is reexamined. To explain the performance limitations of QHM methods, the concept of model order is discussed. Based on this, a high-order sinusoidal modeling method (High-order SM), which can also be viewed as a high-order extension of QHM, is proposed to improve the accuracy of parameter estimation and signal reconstruction in the presence of frequency deviations or noise. The main contributions of this work are summarized as follows:
	\begin{itemize} 
		\item [ 1)] The QHM is revisited from the perspective of frequency mismatches, revealing that its ability to correct such mismatches originates from the structure of its complex time-varying amplitude model. Specifically, the amplitude model in QHM can be interpreted as the first-order Taylor approximation of the frequency mismatch term. However, being restricted to the first order, QHM lacks sufficient expressiveness to accurately model larger frequency mismatches, which fundamentally limits its correction capability.
		\item [ 2)] We are motivated to generalize the QHM to a high-order polynomial form, showing that a higher order yields a more accurate approximation of the mismatch term. However, the lack of sinusoidal structure limits its overall performance. Accordingly, a high-order SM is carefully designed and proposed. The quantitative analysis indicates that the parameter estimation error by the high-order SM exhibits a decrease with increasing model order. Thus, the amplitude, frequency mismatch, and phase can be more precisely estimated, enabling more accurate frequency correction.
		\item [ 3)] Based on high-order SM, the associated optimization problem becomes a nonlinear least-squares problem, for which conventional linear solvers are no longer applicable. We therefore adopt the Levenberg-Marquardt (LM) algorithm for parameter estimation. 
		To enable efficient parallel optimization across multiple frames, we explicitly derive the Jacobian and elaborate the algorithmic steps, thereby avoiding the computational burden introduced by automatic differentiation along undesired dimensions.
	\end{itemize} 

	The structure of this paper is organized as follows. Section II reviews the QHM methods and discusses their limitations. Section III introduces the theoretical motivations of the proposed framework, including new definitions and directions for improvement. In Section IV, the implementation of the proposed high-order model is detailedly elaborated, accompanied by exhaustive pseudocode. Section V conducts quantitative comparisons between the proposed method and existing approaches (QHM and eaQHM) using simulation signals. Section VI demonstrates the application to speech modeling, including both analysis and synthesis. Eventually, Section VII concludes the paper.

	\section{Review of QHM methods}
	\label{S:2}
	This study focuses on signals of the form in Eq. (\ref{eq1}), composed of well-separated sinusoidal components (WSSCs). The WSSCs are assumed to satisfy the following condition:
	\newtheorem{conditions}{Hypothesis} 
	\begin{conditions}\vspace{-0.2cm}
		{\rm \bf (Well-Separated Sinusoidal Components)}  
		Let $\mathcal{B}_{\epsilon,d}$ denote the class of multicomponent signals $x(t) = \sum_{k=1}^{K} A_k(t) \cos(\varphi_k(t))$, such that for all $k$:
		\begin{itemize}
			\item $A_k \in \mathcal{C}^1(\mathbb{R}) \cap L^{\infty}(\mathbb{R})$, with $A_k(t) > 0$ for all $t$;
			\item $\varphi_k \in \mathcal{C}^2(\mathbb{R})$, with $\varphi_k'(t) > 0$ and $\sup_t \varphi_k'(t) < \infty$.
		\end{itemize}
		The instantaneous frequencies are assumed to be well separated, i.e.,
		$|\varphi_{k+1}'(t) - \varphi_k'(t)| \ge 2d, \quad \forall t \in \mathbb{R},\ \forall k \in \{1, \dots, K-1\},$
		where $d$ is the separation parameter.
	\end{conditions}\vspace{-0.2cm}
	Hypothesis 1 restricts the signals to a form in which each component occupies its own region in the TF domain, with sufficient separation ensured by gaps (often referred to as ``no man's land'') \cite{2011Daubechies}. Such  well-separated structures are frequently encountered in the real-world signals, including voiced segments in speech signals or periodic structures in mechanical vibration signals.

	\subsection{Quasi-Harmonic Modeling}
	To overcome the limitation of SM that the frequency is locally inaccurate and stationary, studies, such as \cite{Stylianou2002applying}, \cite{Stylianou1998on}, \cite{stylianou1996harmonic}, explored the use and performance of time-varying models in harmonic modeling. Subsequently, the quasi-harmonic modeling (QHM) incorporates a complex slope term to improve the expressiveness of the model in matching the target signal, as shown as  
	\begin{align}
	x(t)=\left[ \sum\limits_{k=-K}^{K}{\left( {{a}_{k}}+t{{b}_{k}} \right){{e}^{i2\pi {{{\hat{f}}}_{k}}t}}} \right]w(t),\text{   }t\in \left[ -{{T}_{l}},{{T}_{l}} \right],
	\label{eq2}
	\end{align}
	where $\hat{f}_k$ is initial frequency and $a_k, b_k \in \mathbb{C} $ are complex amplitude and slope, respectively. $w(\cdot)$ is the window with length $2T_l$. The term $t b_k$ introduces a complex linear modulation that enables the model to approximate subtle frequency mismatches via the first-order Taylor expansion in the frequency domain, thereby enhancing its frequency correction capability \cite{pantazis2008properties}. The Fourier transform of Eq. (\ref{eq2}) reads
	\begin{align}
	X(f) = \sum_{k=-K}^{K} \left[ a_k W(f - \hat{f}_k) + i \frac{b_k}{2\pi}W'(f - \hat{f}_k) \right],
	\label{eq3}
	\end{align}
	where $W$ and $W'$ denote the Fourier transform of $w(t)$ and its derivative with respect to $f$, respectively. Focusing on the $k$-th component, \cite{pantazis2010iterative} proves that it can be approximated as
	\begin{align}
	X_k(f) &= \scalebox{0.99}{$a_k W(f - \hat{f}_k) + i \frac{b_k}{2\pi}W'(f - \hat{f}_k)$} \notag \\
	&= \scalebox{0.99}{$a_k \left[ W(f - \hat{f}_k) - \frac{\rho^o_{k}}{2\pi} W'(f - \hat{f}_k) + i\frac{\rho^c_{k}}{2\pi} W'(f - \hat{f}_k) \right]$} \notag \\
	&\approx \scalebox{0.99}{$a_k \left[W\left(f - \hat{f}_k - \frac{\rho^o_{k}}{2\pi} \right) + i\frac{\rho^c_{k}}{2\pi} W'(f - \hat{f}_k)\right],$}
	\label{eq4}
	\end{align}
	where $b_k$ is decomposed into parts collinear and orthogonal to $a_k$, i.e., $b_k = \rho^c_{k}a_k + \rho^o_{k}i a_k$. The last approximation is obtained by the first-order Taylor expansion of $W(f - \hat{f}_k - \frac{\rho^o_{k}}{2\pi})$. Applying the inverse Fourier transform to Eq. (\ref{eq4}), we have
	\begin{align}
	x_k(t) \approx a_k \left[e^{i(2\pi \hat{f}_k + \rho^o_{k})t} + \rho^c_{k}te^{i2\pi \hat{f}_kt}\right],
	\label{eq5}
	\end{align} 
	which indicates that the collinear coefficient $\rho^c_{k}$ accounts for an amplitude slope and the $k$-th frequency deviations (denoted as $\eta_k$) can be estimated by the orthogonal coefficient $\rho^o_{k}$, as
	\begin{align}
	\eta_k = f_k - \hat{f}_k = \frac{\rho^o_{k}}{2\pi}=\frac{ \langle ia_k, b_k \rangle}{2\pi|a_k|^2}=\frac{a^{R}_kb^{I}_k - a^{I}_kb^{R}_k}{2\pi|a_k|^2}. 
	\label{eq6}
	\end{align}
	where $\langle \cdot, \cdot \rangle$ denotes the complex inner product, and $a^{R}_k, a^{I}_k$ and $b^{R}_k,b^{I}_k$ are the real and imaginary parts of $a_k,b_k$, respectively. Once $a_k$ and $b_k$ are obtained through LS optimization, the frequency mismatch, amplitude, and phase at frame center can be determined. Nevertheless, if the initial frequency mismatch is large, an iterative correction \cite{pantazis2010iterative} is necessary. Unfortunately, as such iteration still presumes intra-frame stationarity, estimation errors persist and cannot be fully eliminated.

	To accommodate time-varying frequency components, aQHM \cite{pantazis2008properties} substitutes the exponential term in Eq.~(\ref{eq2}) with a non-stationary phase function, as follows:
	\begin{align}
	& x(t)=\left\{ \sum\limits_{k=-K}^{K}{\left( {{a}_{k}}+t{{b}_{k}} \right){{e}^{i\left[ {{{\hat{\varphi }}}_{k}}(t+{{t}_{l}})-{{{\hat{\varphi }}}_{k}}({{t}_{l}}) \right]}}} \right\}w(t), \notag \\ 
	& \text{ }t\in \left[ -{{T}_{l}},{{T}_{l}} \right]  
	\label{eq7}
	\end{align}
	where $\hat{\varphi}_{k}$ is the phase function of the $k$-th component and $t_l$ denotes the center of the $l$-th frame ($l=1,\cdots,L$). aQHM also adopts the iterative correction strategy, in which the phase is updated at each iteration from the previously estimated frequency $\hat{f}_k^{\text{new}}$ using the following expression:
	\begin{align}
	{\varphi }_k(t)=\varphi_k(t_l) + \int_{{t_l}}^{t_l+t}{2\pi \hat{f}_k^{\text{new}}(u)}\text{d}u, \text{ }t\in [ -{{T}_{l}},{{T}_{l}}].
	\label{eq8}
	\end{align}
	This updated phase is then used as the initial condition for LS to obtain more accurate $a_k$ and $b_k$, further estimating the current frequency mismatch. Through multiple iterations, the estimated frequency converges more rapidly and accurately toward the ground truth, also contributing to more accurate signal reconstruction.

	For certain signals with strong amplitude modulation, QHM and aQHM still fail to accurately estimate the parameters, particularly the amplitude. To address this limitation, eaQHM extends aQHM by introducing an adaptive amplitude amplifier, as defined below: 
	\begin{align}
	& x(t)=\left\{ \sum\limits_{k=-K}^{K}{\left( {{a}_{k}}+t{{b}_{k}} \right)\frac{{{A}_{k}}(t+{{t}_{l}})}{{{A}_{k}}({{t}_{l}})}{{e}^{i\left[ {{{\hat{\varphi }}}_{k}}(t+{{t}_{l}})-{{{\hat{\varphi }}}_{k}}({{t}_{l}}) \right]}}} \right\}w(t), \notag \\ 
	& \text{ }t\in \left[ -{{T}_{l}},{{T}_{l}} \right]  
	\label{eq9}
	\end{align}
	\noindent where $A_k(t)$ is the amplitude of the $k$-th component. Such an amplifier is derived from the ratio between the instantaneous amplitude over the entire frame and that at the frame center, leading the model to more effectively track rapid amplitude modulation within a frame. Likewise, eaQHM adopts an iterative algorithm, where the amplitude and phase estimated from the previous iteration are considered as initial values for the current iteration. LS optimization is then applied to obtain more accurate parameters, enabling both amplitude and frequency to be more rapidly and precisely determined.
	
	\subsection{Limitations of QHM methods}   
	Although QHM methods exhibit strong adaptability to modulated signals, some limitations still exist.
	\subsubsection{Insufficient Parameter Estimation}
	While QHM accurately estimates the amplitudes, phase, and frequency deviations when the initial frequency error is small, its performance degrades as the error increases. In such cases, the estimation of $a_k$ and $b_k$ deteriorates, leading to inaccurate or even amplified frequency deviations and amplitudes. aQHM and eaQHM successively introduce nonlinear structures into the model's amplitude and phase parts for higher adaptability in parameter estimation, and adopt iterative schemes to progressively refine the frequency across iterations. However, when the initial frequency error is large, the estimation accuracy is still limited. Moreover, the iterations may amplify the error instead of reducing it, especially in multicomponent signals where closely spaced components aggravate the problem.
	
\begin{figure}[!t]\centering
		\includegraphics[width=8.5cm]{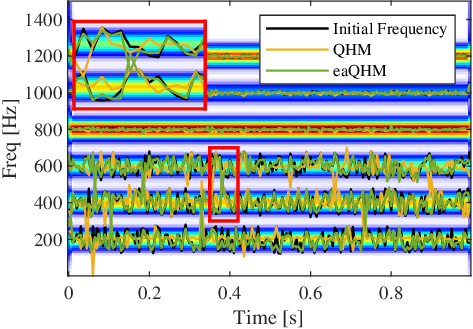}
		\vspace{-0.3cm}
		\caption{Frequency estimates obtained by QHM and eaQHM.}
		\label{FIG1}
		\vspace{-0.5cm}
	\end{figure}

\begin{figure}[!t]\centering
		\includegraphics[width=8.89cm]{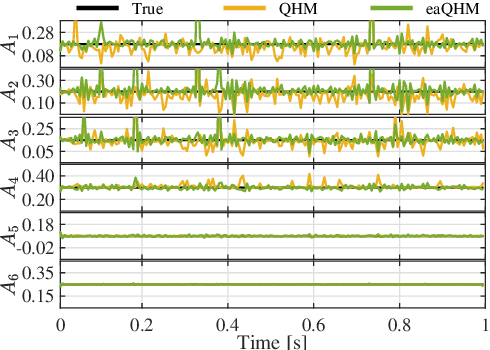}
		\vspace{-0.7cm}
		\caption{Amplitude estimates obtained by QHM and eaQHM.}
		\label{FIG2}
		\vspace{-0.5cm}
	\end{figure}

	To visualize this issue, a simulated signal with six WSSCs,
	\begin{align}
	x(t) = \sum_{k=1}^{6}\scalebox{0.925}{$A_ke^{i\left(400 \pi  k t + \phi_k\right)},  A_k \sim \mathcal{U}[0, 1], \phi_k \sim \mathcal{U}[-\pi, \pi],$}
	\label{eq10}
	\end{align}
	where $\phi_k$ is the $k$-th initial phase, is employed to be analyzed by QHM and eaQHM. The initial frequencies of the first three components are corrupted by large random deviations ($\eta_k  \in [-80,80]$), whereas those of the last three receive smaller deviations ($\eta_k  \in [-10,10]$). The frequency estimates and the spectrogram of the signal are demonstrated in Fig. \ref{FIG1}. Evidently, QHM and eaQHM effectively depict the accurate frequency trajectories for components with modest initial frequency mismatches, whereas their performance and robustness significantly diminish under substantial mismatches, resulting in the oscillating frequency trajectories of the first three components. Figure ~\ref{FIG2} illustrates the amplitude estimates obtained by QHM and eaQHM, along with the ground truth. Similarly, components with small initial frequency mismatches yield more accurate amplitude estimates, whereas larger initial mismatches lead to pronounced amplitude estimation errors.
	
	Notably, around $t=0.4s$, the estimated frequencies of the second and third components by eaQHM are erroneously swapped, i.e., the second component's frequency is assigned the region of the third, and vice versa, as shown by the enlarged view in Fig. \ref{FIG1}. This occurs due to large initial frequency mismatches that fall outside the effective correction range of QHM.
	The iterative steps in eaQHM may gradually and mistakenly amplify the frequency mismatch. Furthermore, since the iteration in aQHM and eaQHM relies on interpolated frequencies to obtain phase across frames, large frequency mismatch may affect adjacent frames, deteriorating the estimation accuracies.

	\subsubsection{Heavy computation}
	Although QHM estimates parameters efficiently via LS, aQHM and eaQHM rely on iterative procedures that incur substantial computational burden. Each iteration requires reconstructing the signal, involving repeated cubic spline interpolation for frequency and linear interpolation for amplitude. The spline interpolation and LS optimization, in particular, are time-consuming and significantly increase the computational cost. Moreover, to ensure sufficient adaption of the model to the target signals, these methods typically employ the signal-to-reconstruction-error ratio (SRER) as a stopping criterion, terminating the iterations once SRER stops increasing. Otherwise, the process continues, potentially incurring high computational costs. In the case of large initial frequency deviations, more iterations are required to achieve acceptable accuracy. Conversely, restricting the number of iterations may degrade performance, thus highlighting the trade-off between estimation accuracy and computational efficiency.

	\section{High Order Approximation}
	\label{S:3} 
	To overcome the limitations of QHM methods, we revisit the model from the perspective of frequency mismatch and give the explanation of the degradation of QHM in mismatch correction, i.e., the limited order of amplitude. To address this, we are inspired to increase the order and give the definition of the high-order sinusoidal model, significantly improving the parameter extraction accuracy.
	
	\vspace{-0.4cm}
	\subsection{On the Insufficiency of Low-Order Models in QHM}
	Here we revisit the QHM methods from the perspective of frequency mismatch. Considering a stationary complex multicomponent signal, it can be expressed as 
	\begin{align}
		x(t) = \sum_{k=-K}^{K}A_k e^{i\left(2\pi f_kt + \phi_k \right)},
		\label{eq11}
	\end{align}
	where $A_k$, $f_k$, and $\phi_k$ are the amplitude, frequency and initial phase of $k$-th component. Assuming that the truth frequency $f_k$ deviates from an observed value $\hat{f}_k$ by a mismatch $\eta_k$, i.e., $f_k = \hat{f}_k + \eta_k$, the $k$-th component reads
	\begin{align}
	x_k(t) &= A_k e^{i\left(2\pi f_kt + \phi_k \right)}	\notag \\
	       &= A_k e^{i\phi_k}e^{i2\pi\eta_k t}e^{i2\pi\hat{f}_k t}.
	\label{eq12}
	\end{align}
	Applying the first-order Taylor expansion to $e^{i2\pi\eta_k t}$ around $t=0$ yields:
	\begin{align}
		x_k(t) &\approx A_k e^{i\phi_k}\left(1 + i2\pi \eta_k t\right) e^{i2\pi\hat{f}_k t} \notag \\
			   & = \left(a_k + tb_k\right)e^{i2\pi\hat{f}_k t},
		\label{eq13}
	\end{align}
	where $a_k = A_k e^{i\phi_k}$ and $b_k = A_k e^{i\phi_k} i 2\pi \eta_k $ are considered. Obviously, Eq. (\ref{eq13}) is identical to the $k$-th component of QHM in Eq. (\ref{eq2}), revealing that QHM captures frequency mismatches using only the first-order model. Although once $a_k$ and $b_k$ are obtained, the amplitude and frequency mismatch can be easily estimated by
	\begin{subequations}
		\begin{align}\label{eq14}
			\eta_k &\approx  \Re \left\{ \frac{b_k}{i2\pi a_k} \right\} =  \frac{a_k^Rb_k^I-a_k^Ib_k^R }{2\pi |a_k|^2}  \\
			A_k &\approx |a_k| = \sqrt{{a_k^R}^2 + {a_k^I}^2},
		\end{align} 
	\end{subequations}
	this approximation still introduces residual error due to model limitations and truncation of higher-order terms. This motivates the adoption of higher-order expressions to enhance the model's expressive capacity, enabling more accurate approximation of the mismatch term.
	
	\subsection{Towards High-Order Model}
	
	\newtheorem{myDef}{Definition}
	 To investigate the impact of higher-order expressions on estimation accuracy, a natural way is to increase the amplitude polynomial order. Thus, we begin with the definition of a higher-order polynomial model for a monocomponent signal.	
	\begin{myDef}
		{\rm \bf (Higher-Order Polynomial Model)} For a single sinusoid, $x(t) \in L^2(\mathbb{R})$, with true frequency $f$, its high-order polynomial model within $t\in \left[ -{{T}_{l}},{{T}_{l}} \right]$ is defined as
		\begin{align}
		x_\text{PM}^{(N)}(t) = \left[ \left( \sum_{n=0}^{N} c_{n} t^n \right) e^{i 2\pi \hat{f} t} \right]w(t), \quad c_{n} \in \mathbb{C},
		\label{eq15}
		\end{align}
		where $w(\cdot)$ is an analysis window, and $\hat{f}$ is the observed frequency that differs from $f$ by a small deviation $\eta = f - \hat{f}$. The term $\sum_{n=0}^{N} c_n t^n$ is referred to as the $N$th-order polynomial amplitude, characterizing local amplitude structure and implicit frequency mismatch of the sinusoid.
		
	\end{myDef}
	
	{\rm \bf Remark:} This model uses $N\!+\!1$ independent coefficients $\{c_n\}_{n=0}^{N}$ to capture local parameters (amplitude, phase, and frequency mismatch), which are estimated by fitting the model to the framewise signal \cite{Stylianou1998on}, \cite{stylianou1996harmonic}.
	Although such a higher-order polynomial model is more expressive, its lack of sinusoidal structure introduces several fundamental limitations that affect parameter estimation, as discussed below:
	\begin{enumerate}
		\item Considering the $N$th-order Taylor series of mismatch,
		\begin{align}
		Ae^{i\phi}e^{i2\pi\eta t} \approx Ae^{i\phi}\sum_{n=0}^{N}\frac{(i2\pi\eta t)^n}{n!}= \sum_{n=0}^{N} c_{n} t^n,
		\label{eq16}
		\end{align}
		the coefficients $\{c_n\}_{n=1}^N$ are theoretically related to the frequency mismatch $\eta$, and thus mutually dependent, i.e.,
		\begin{align}
		\eta^n = \frac{n! c_n}{(i2\pi)^n c_0}, \quad c_{n} = \frac{i 2\pi\eta}{n} c_{n-1}.
		\label{eq17}
		\end{align}
		However, such relations generally fail to hold in practical estimation.
		
		\item This model is a general polynomial with $N\!+\!1$ free coefficients and lacks strong structural correspondence with the exponential mismatch term $e^{i2\pi \eta t}$. 
		In the presence of noise, the fitted polynomial can easily diverge from $e^{i2\pi \eta t}$, leading to non-robust parameter estimates.
	\end{enumerate}

	To better capture the sinusoidal structure of $e^{i2\pi \eta t}$, we are inspired to reformulate the model to achieve the strong structural alignment with target. Observing Eq. (\ref{eq17}), one can derive the following relationship among the coefficients:
	\begin{align}
	c_2 = \frac{c_1^2}{2c_0},c_3 = \frac{c_1^3}{3!c_0^2},\cdots,c_N = \frac{c_1^N}{N!c_0^{N-1}}.
	\label{eq18}
	\end{align}
	Substituting Eq. (\ref{eq18}) into Eq. (\ref{eq15}), we have
	\begin{align}
	x(t) = (c_0 + c_1t + \frac{c_1^2}{2c_0}t^2 + \cdots +  \frac{c_1^N}{N!c_0^{N-1}}t^N)e^{i 2\pi \hat{f} t}w(t).
	\label{eq19}
	\end{align}
    In this way, the Eq. (\ref{eq17}) always holds. 
    Specifically, $\{c_n\}_{n=1}^{N}$ are strictly derived by the Taylor expansion of the mismatch term, compressing the $N\!+\!1$ coefficients into only two free coefficients $(c_0, c_1)$. This enforces structural alignment with $e^{i2\pi \eta t}$.
    To avoid placing the parameters in the denominator of the polynomial, which may cause numerical instability in practical computation, we simplify $\{c_n\}_{n=0}^{N}$ by $a$ and $b$, as 
    \begin{align}
    c_0 = a,\quad \frac{c_1^n}{c_0^{n}} = b^n.
    \label{eq19_2}
    \end{align}
    We propose a specific high-order model, referred to as the \emph{High-Order Sinusoidal Model} (high-order SM), and extend it to multicomponent signals. Its formal definition is given below.
	\begin{myDef}
		{\rm \bf (Higher-Order Sinusoidal Model)} For a multicomponent signal, $x(t) \in L^2(\mathbb{R})$, with observed frequencies $\{\hat{f}_k\}_{k=-K}^{K}$, its high-order sinusoidal model for recovering amplitudes and frequency mismatches is defined as
		\begin{align}
		&x^{(N)}(t) = \left[\sum_{k=-K}^{K}a_k \left( \sum_{n=0}^{N}\frac{b_k^nt^n}{n!}\right)e^{i 2\pi \hat{f}_k t} \right] w(t),\notag \\
		&a_k, b_k \in \mathbb{C}, \quad t\in \left[ -{{T}_{l}},{{T}_{l}} \right]
		\label{eq20}
		\end{align}
		where $w(\cdot)$ is the analysis window and $N$ is the order. $a_k$ and $b_k$ are used to estimate the complex amplitudes $\hat{A}_ke^{i\hat{\phi}_k}$ and frequency mismatches $\hat{\eta}_k$ by 
		\begin{align}
		\hat{A}_k = |a_k|,\hat{\phi}_k = \angle a_k , \hat{\eta}_k = \frac{\Im \{b_k\}}{2\pi}.
		\label{eq21}
		\end{align}
	\end{myDef}

	{\rm \bf Remark:} This model significantly improves the structural consistency with the sinusoidal form, such that its representation remains a valid Taylor expansion of the complex exponential even as the model order increases. Furthermore, it reduces the parameter space to only two key coefficients: $a_k$	(complex amplitude) and $b_k$ (frequency mismatch), leading to a more compact and stable estimation framework.

\begin{figure}[!t]\centering
		\includegraphics[width=8.5cm]{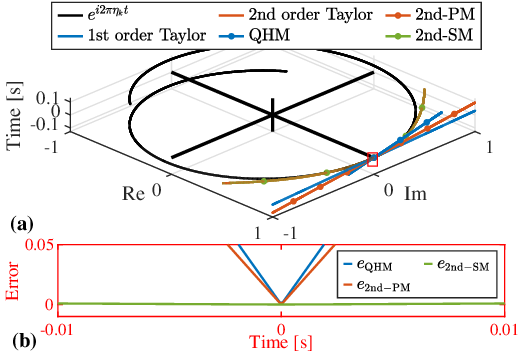}
		\vspace{-0.3cm}
		\caption{(a) Visualization of the mismatch term (black), its 1st- and 2nd-order Taylor expansions (solid), and the corresponding approximations obtained by QHM (blue line with markers, compared to 1st-order Taylor), 2nd-PM, and 2nd-SM (orange and green lines with markers, compared to 2nd-order Taylor).  
			(b) The Euclidean distance between each method and its corresponding Taylor expansion, showing the deviation of the complex trajectories in [-0.01,0.01].
		}
		\label{FIG3}
		\vspace{-0.3cm}
	\end{figure}

	To visualize the effectiveness of high-order SM, Fig. \ref{FIG3}(a) depicts the curve of the mismatch term $e^{i2\pi\eta_k t}$ when $\eta_k = 5 Hz$, along with its Taylor expansions and the estimation results from QHM and second-order polynomial model (2nd-PM) and second-order SM (2nd-SM). Evidently, the closer the model curve aligns with its corresponding Taylor expansion of the same order, the more accurately the frequency mismatch $\eta_k$ can be estimated, i.e., $e^{b_kt}\rightarrow e^{i2\pi\eta_k t}$. QHM significantly diverges from the 1st-order Taylor expansion while 2nd-PM also exhibits a noticeable deviation from the 2nd-order Taylor expansion, reflecting their insufficient ability to approximate the frequency mismatch structure. In contrast, the 2nd-SM almost perfectly overlaps with the 2nd-order Taylor expansion. In Fig. \ref{FIG3}(b), the Euclidean distances between each method and its corresponding Taylor expansion
	are given, i.e., (QHM vs. 1st-order Taylor) and (2nd-PM/2nd-SM vs. 2nd-order Taylor), demonstrating the superior capability of 2nd-SM in capturing the frequency mismatch structure and estimating the complex amplitude and frequency parameters more accurately.
	
\begin{figure}[!t]\centering
		\includegraphics[width=8.5cm]{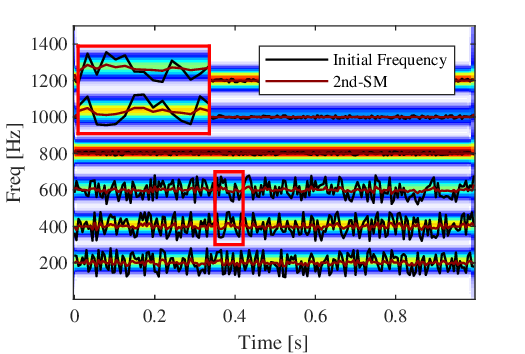}
		\vspace{-0.3cm}
		\caption{Frequency estimates obtained by 2nd-SM.}
		\label{FIG4}
		\vspace{-0.3cm}
	\end{figure}

\begin{figure}[!t]\centering
		\includegraphics[width=8.89cm]{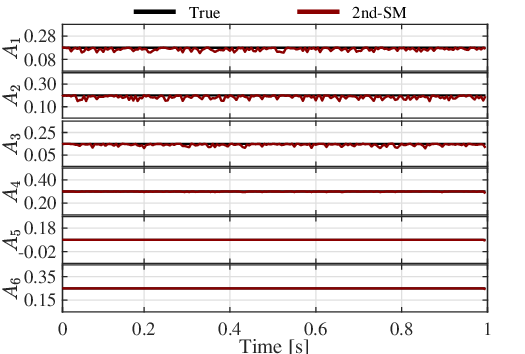}
		\vspace{-0.7cm}
		\caption{Amplitude estimates obtained by 2nd-SM.}
		\label{FIG5}
		\vspace{-0.5cm}
	\end{figure}

	To show the improvement brought by high-order SM, we adopt a 2nd-SM to make the improved versions of Figs. \ref{FIG1} and \ref{FIG2}, given in Figs. \ref{FIG4} and \ref{FIG5}, respectively. Compared with QHM and eaQHM, the 2nd-SM yields noticeably more reliable frequency estimates, even under substantial initial frequency mismatch. In addition, it provides significantly improved amplitude estimation
	, as shown in Fig. \ref{FIG5}.
	

	 \subsection{Quantitative Performance Analysis of High-Order SM in Parameter Estimation}
\begin{figure}[!t]\centering
		\includegraphics[width=8.89cm]{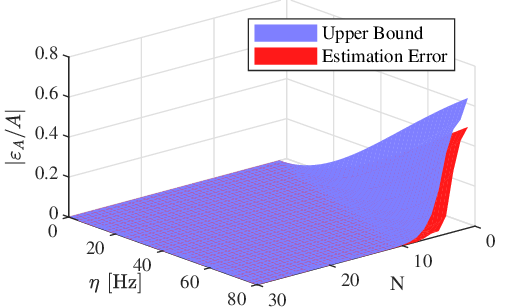}
		\vspace{-0.7cm}
		\caption{Relative estimation errors of amplitude $A$ as a function of the initial frequency mismatch $\eta$ and the model order $N$ ($T=10$ms).}
		\label{FIG6}
	\end{figure}

\begin{figure}[!t]\centering
		\includegraphics[width=8.89cm]{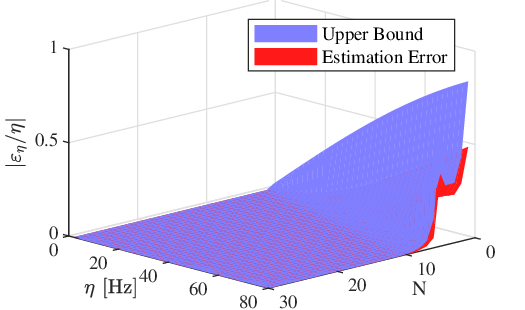}
		\vspace{-0.7cm}
		\caption{Relative estimation errors of frequency mismatch $\eta$ as a function of the initial frequency mismatch $\eta$ and the model order $N$ ($T=10$ms).}
		\label{FIG7}
		\vspace{-0.5cm}
	\end{figure}

	 In this subsection, we quantitatively analyze the accuracy of high-order SM in parameter estimation, i.e., amplitude and frequency mismatch. We focus on a monocomponent signal
	 \begin{align}
	 x(t) = A e^{i\left(2\pi ft + \phi \right)} w(t) = Ae^{i\phi} e^{i2\pi \eta t}e^{i2\pi \hat{f} t }w(t)
	 \label{eq22}
	 \end{align}
	 where $\hat{f}$ and $\eta$ are initially estimated frequency and mismatch, respectively. $w(t)$ is the window.
	 Considering that the estimation is performed by minimizing the least-squares error within a frame $t\in[-T,T]$, we have the following theorem giving the estimates and biases of the amplitude $A$ and frequency mismatch $\eta$, showing that the proposed model achieves higher accuracy in parameter estimation as the model order increases.
	\begin{theorem}
		Let $\hat{a}$ and $\hat{b}$ be the global optima of the following least-squares problem within the frame $t\in \left[ -{{T}},{{T}} \right]$:
		\begin{align}
		(\hat{a}, \hat{b}) = \arg\min_{a, b} \frac{1}{2}\left\| x^{(N)}(t; a, b)- x(t) \right\|^2,
		\label{eq23}
		\end{align}
		where $ x^{(N)}(t; a, b )$ is the $N$th-order sinusoidal model. The amplitude $A$ and frequency mismatch $\eta$ can be estimated from $\hat{a}$ and $\hat{b}$ as
		\begin{align}\label{eq24}
		\hat{A} &= \frac{\left| \langle x(t), y^{(N)}(t; b) \rangle\right|}{\left\|y^{(N)}(t; b)\right\|^2}, \\
		\hat{\eta} &= \Im \left\{\arg\max_{b} \frac{|\langle x(t), y^{(N)}(t; b) \rangle|^2}{\left\|y^{(N)}(t; b)\right\|^2} \right\} /2\pi
		\label{eq25}
		\end{align}
		where $y^{(N)}(t; b) = w(t)e^{i2\pi \hat{f}t}\sum_{n=0}^{N}\frac{(bt)^n}{n!}$ are complex bases.
	\end{theorem}
	\noindent The proof of Theorem 1 is given in Appendix \ref{app:theorem1}. The following presents an analysis of the estimation accuracy for the amplitude and frequency mismatch.
	\begin{prop}
		Let $\varepsilon_A$ and $\varepsilon_\eta$ denote the estimation errors of the amplitude and frequency mismatch obtained by the $N$-th order SM. There exist constants $C_2,C_3 \in \mathbb{R^+}$ such that the estimation errors satisfy:
		\begin{align}\label{eq26}
		\left|\varepsilon_A\right| &\leq \frac{A|2\pi\eta|^{N+1}}{C_2(N+1)!}\sqrt{\int_{-T}^{T}t^{2(N+1)}w^2(t)dt}, \\
		\left|\varepsilon_\eta\right| &\leq \frac{|2\pi|^{N}|\eta|^{N+1}}{C_3(N+1)!}\sqrt{\int_{-T}^{T}t^{2(N+1)}w^2(t)dt}.
		\label{eq27}
		\end{align}
	\end{prop}
	\noindent The proof of Proposition 1 is shown in Appendix \ref{app:proposition1}, which is based on the proof of Theorem 1. Proposition~1 indicates that the estimation errors are determined by both the initial frequency mismatch $\eta$ and the model order $N$. In general, a larger initial frequency mismatch leads to a larger estimation error. Nevertheless, Eqs. (\ref{eq26}) and (\ref{eq27}) also reveal that, for sufficiently large $N$, the estimation error converges to zero. 
	To provide a more intuitive illustration, Figs.~\ref{FIG6} and \ref{FIG7} depict the values and corresponding upper bounds of relative estimation errors for the amplitude $A$ and initial frequency mismatch $\eta$ as functions of $\eta$ and the model order $N$, with Hamming window and $T = 10$ ms, respectively. Evidently, increasing the $N$ leads to a substantial reduction in the estimation errors for both the amplitude and the frequency mismatch.


	\section{Implementation of High-Order SM}
	
	In this section, the implementation of the high-order SM is focused, in terms of the choice of the optimization method and the algorithmic steps of optimization and reconstruction. 
	
	\vspace{-0.3cm}
	\subsection{Optimization method}
	First, we focus on the choice of the optimization method. As shown in Eq. (\ref{eq23}), the problem is formulated as a LS estimation of $a$ and $b$. Unlike the QHM methods, where all parameters are linear and can be directly estimated via projection \cite{pantazis2008properties}, high-order SM is linear in $a$ but nonlinear in $b$, rendering direct LS method inapplicable. An appropriate nonlinear optimization strategy is essentially required.
	
	A straightforward approach is to employ gradient descent (GD) with a nonadaptive step size $\alpha$ to iteratively approximate the solution, as shown as 
	\begin{align}
		\mathbf{p}_{j+1} = \mathbf{p}_j - \alpha \mathbf{J}^\dagger(\mathbf{p}_j)\mathbf{r}(\mathbf{p}_j),
		\label{eq28}
	\end{align}
	where $(\cdot)^\dagger$ is the Hermitian transpose. $\mathbf{p}=[p_1,\cdots, p_M]^\top$ is the parameter vector, and $\mathbf{r}=[r_1,\cdots, r_D]^\top$ is the residual vector, i.e., $\mathbf{r} =  x^{(N)}(t_d; a, b)- x(t_d)$ where $t_d$ is the discrete time points in each frame. $j$ indexes the iterations and $\mathbf{J}(\mathbf{p}_j)$ is the Jacobian in the $j$-th iteration, which is calculated by 
	\begin{align}
		\mathbf{J}(\mathbf{p}_j) = \left[ \frac{\partial r_d}{\partial p_{j,m} } \right], \quad m = 1,\cdots,M; d = 1,\cdots, D,
		\label{eq29}
	\end{align}
	where $M$ and $D$ are the dimensions of $\mathbf{p}$ and $\mathbf{r}$, respectively.
	While stable, GD suffers from slow convergence due to its nonadaptive step size causing the oscillation near the optima \cite{bpqhm}. In contrast, Newton's method adaptively adjusts the step size using second-order derivatives:
	\begin{align}
		\mathbf{p}_{j+1} = \mathbf{p}_j - \mathbf{H}^{-1}(\mathbf{p}_j)\mathbf{J}^\dagger(\mathbf{p}_j)\mathbf{r}(\mathbf{p}_j),
		\label{eq30}
	\end{align} 
	where $\mathbf{H}(\mathbf{p}_j)$ is the Hessian, calculated by
	\begin{align}
	\scalebox{1.0}{$\mathbf{H}(\mathbf{p}_j) 
	= \left[ \sum_{d=1}^{D} 
	\left(
	\frac{\partial r_d}{\partial p^*_{j,m}} \frac{\partial r^*_d}{\partial p_{j,q}}
	+ r^*_d \frac{\partial^2 r_d}{\partial p^*_{j,m} \partial p_{j,q}}
	\right)
	\right]_{m,q=1}^{M}.$}
	\label{eq31}
	\end{align}
	The Gauss-Newton method (GNM) replaces $\mathbf{H}(\mathbf{p}_j)$ by $ \mathbf{J}^\dagger(\mathbf{p}_j) \mathbf{J}(\mathbf{p}_j)$ to reduce the computational cost. Nevertheless, both Newton's method and GNM can be unstable or divergent when $\mathbf{H}(\mathbf{p}_j)$ or $ \mathbf{J}^\dagger(\mathbf{p}_j) \mathbf{J}(\mathbf{p}_j)$ are non-positive definite, especially far from the optima. 
	
	The Levenberg-Marquardt (LM) algorithm introduces a scaled identity matrix $\lambda_j \mathbf{I}$ ($\lambda_j > 0$ ) to $ \mathbf{J}^\dagger(\mathbf{p}_j) \mathbf{J}(\mathbf{p}_j)$, as
	\begin{align}
	\mathbf{p}_{j+1} = \mathbf{p}_j - \left[\mathbf{J}^\dagger(\mathbf{p}_j) \mathbf{J}(\mathbf{p}_j) + \lambda_j \mathbf{I} \right]^{-1}\mathbf{J}^\dagger(\mathbf{p}_j)\mathbf{r}(\mathbf{p}_j),
	\label{eq32}
	\end{align} 
	where $\lambda$ is the damping factor, which ensures positive definiteness and improves numerical stability while retaining fast convergence. LM effectively combines the advantages of GD and GNM by adjusting $\lambda$: a large $\lambda$ makes the method behave more like GD, which is preferable in the early stages to maintain stability; a small $\lambda$ makes it closer to GNM, avoiding oscillations and accelerating convergence near the optimum. For our problem in Eq. (\ref{eq23}), LM serves as an effective and robust choice to jointly estimate $a$ and $b$. Next, the details of signal modeling using high-order SM with LM are presented.
	
	In practical implementations, the signal is processed on a frame-by-frame basis. To improve computational efficiency, a parallel scheme is adopted to simultaneously estimate the parameters of all frames. For clarity, the following derivations are based on the multicomponent signal in Eq.~(\ref{eq11}) within the $l$-th frame ($l=1,\cdots,L$), and all variables in this subsection are implicitly indexed by $l$. In the first step, the initial frequencies of all components, denoted as $\mathbf{\hat{f}} = [\hat{f}_{-K} ,\cdots,\hat{f}_{K} ]^\top$, should be roughly detected from the TF representation, while the hyper-parameter $N$, $\lambda$, and $T$ are initialized. Second, the high-order SM is solved by LM. Let $\mathbf{p} _j$ denote the parameter vector composed of $a_k$ and $b_k$, as
	\begin{align}
\mathbf{p} _j = \begin{bmatrix} \mathbf{a} _j \\ \mathbf{b} _j \end{bmatrix} 
= \left[ a_{-K} ,\dots,a_{K} ,b_{-K} ,\dots,b_{K}  \right]_j^\top \in \mathbb{C}^{M \times 1},
	\label{eq33}
	\end{align}
	where $M = 4K +2$. Then, the Jacobian is calculated as 
	\setlength{\arraycolsep}{1pt}
	\renewcommand{\arraystretch}{0.5}
	\begin{align}
		\mathbf{J}(\mathbf{p} _j) &=
		\begin{bmatrix}
		\frac{\partial r_1 }{\partial a _{-K}} & \cdots &\frac{\partial r_1 }{\partial a _{k}} & \cdots & \frac{\partial r_1 }{\partial a _K}, & \frac{\partial r_1 }{\partial b _{-K}} & \cdots &\frac{\partial r_1 }{\partial b _{k}} &\cdots & \frac{\partial r_1 }{\partial b _K}\\
		\vdots & \ddots & \vdots & \ddots& \vdots& \vdots & \ddots & \vdots& \ddots& \vdots\\
		\frac{\partial r_d }{\partial a _{-K}} & \cdots &\frac{\partial r_d }{\partial a _{k}} & \cdots & \frac{\partial r_d }{\partial a _K}, & \frac{\partial r_d }{\partial b _{-K}} & \cdots &\frac{\partial r_d }{\partial b _{k}} &\cdots & \frac{\partial r_d }{\partial b _K}\\
		\vdots & \ddots & \vdots & \ddots& \vdots& \vdots & \ddots & \vdots& \ddots& \vdots\\
		\frac{\partial r_D }{\partial a _{-K}} & \cdots &\frac{\partial r_D }{\partial a _{k}} & \cdots & \frac{\partial r_D }{\partial a _K}, & \frac{\partial r_D }{\partial b _{-K}} & \cdots &\frac{\partial r_D }{\partial b _{k}} &\cdots & \frac{\partial r_D }{\partial b _K}
		\end{bmatrix}_j  \notag \\
		&\in \mathbb{C}^{D \times M},
		\label{eq34}
	\end{align}
	\setlength{\arraycolsep}{5pt}
	\renewcommand{\arraystretch}{1}
	\noindent where the following equations are considered
	\begin{align}
	\left[ \frac{\partial r_d }{\partial a _{k}} \right]_j& = w(t_d) e^{i2\pi \hat{f}_k   t_d} \sum_{n=0}^{N}\frac{(b _{j,k}t_d)^n}{n!},
	\label{eq35} \\
	\left[ \frac{\partial r_d }{\partial b _{k}} \right]_j& =a _{j,k} w(t_d) e^{i2\pi \hat{f}_k   t_d} \sum_{n=1}^{N}\frac{b_{j,k}^{n-1}{t_d}^n}{(n-1)!}.
	\label{eq36}
	\end{align}
	Substituting the parameters and Jacobian into Eq. (\ref{eq32}), the parameters will be updated, as
	\begin{align}
	\scalebox{0.94}{$\begin{bmatrix} \mathbf{a} _{j+1} \\ \mathbf{b} _{j+1} \end{bmatrix} = \begin{bmatrix} \mathbf{a} _j \\ \mathbf{b} _j \end{bmatrix} - \left[\mathbf{J}^\dagger(\mathbf{p} _j) \mathbf{J}(\mathbf{p} _j) + \lambda _j \mathbf{I} \right]^{-1}\mathbf{J}^\dagger(\mathbf{p} _j)\mathbf{r}(\mathbf{p} _j)$},
	\label{eq37}
	\end{align} 
	where the residual $	\mathbf{r}(\mathbf{p} _j)$ is calculated by
	\begin{align}
	\mathbf{r}(\mathbf{p} _j) =  [x^{(N)}(t_d; a _{j,k}, b _{j,k})- x(t_d)], d = 1,\cdots,D.
	\label{eq38}
	\end{align} 
	
	To ensure efficient convergence, the algorithm aims to maintain stability when the parameters are far from the optima, while achieving accurate solutions near the optima. Accordingly, the scale factor $\lambda_j $ is adaptively adjusted and the parameters are selectively updated based on the comparison of residual norms at each iteration. Specifically, for the $l$-th frame, the scale factor and parameters are updated as
	\begin{equation}
	\lambda_{j+1}  =
	\begin{cases}
	 \lambda_j  / \gamma, & \text{if } \|\mathbf{r} (\mathbf{p}_{j+1} )\| < \|\mathbf{r} (\mathbf{p}_j )\|, \\[6pt]
	\gamma \, \lambda_j , & \text{otherwise},
	\end{cases}
	\label{eq39}
	\end{equation}
	\begin{equation}
	\mathbf{p}_{j+1}  =
	\begin{cases}
	\mathbf{p}_{j+1} , & \text{if } \|\mathbf{r} (\mathbf{p}_{j+1} )\| < \|\mathbf{r} (\mathbf{p}_j )\|, \\[6pt]
	\mathbf{p}_j , & \text{otherwise},
	\end{cases}
	\label{eq40}
	\end{equation}
	where $\gamma >1 $ (typically set to 10) is a positive coefficient controlling the adjustment rate of $\lambda_j $. The updated parameters are used as the initial values for the next iteration, where the Jacobian and the residuals are recomputed to perform the next update. Repeating this process for $J$ iterations allows the parameters to rapidly converge to their optimal values, from which the amplitude $\hat{A}_k(t_l)$, phase $\hat{\varphi}_k(t_l)$, and frequency $\hat{f}^{\text{final}}_k(t_l)$ mismatch for each frame can be estimated by Eq. (\ref{eq21}). Algorithm \ref{high-order-SM} provides the pseudocode for high-order SM, encapsulating the entire process that has been outlined so far.
	
	\begin{algorithm}[!t]
		\caption{High Order Sinusoidal Modeling}
		\begin{algorithmic}
			\STATE {\textbf{Step 1: Initialization}\\
				Input the signal $x(t)$, the sampling rate $f_s$, the maximum iteration $J$, the frame-shift, window length $2T$, and adjustment coefficient $\gamma$; \\
				Detect the initial frequencies $\mathbf{\hat{f}}$ and initialize the damping factor $\lambda_1 $ and parameter matrix $\mathbf{p} _1$;
			}
			\STATE {\textbf{Step 2: Iterative LM optimization (parallel over frames)}\\
				\STATE Compute the initial residual $\mathbf{r}(\mathbf{p} _1)$ by Eq. (\ref{eq38});\\
				\FOR{$j=1:J$}
				\STATE Compute Jacobians $\mathbf{J}(\mathbf{p} _j)$ for all frames $l=1,\dots,L$;\\
				\STATE Obtain updated parameters $\mathbf{p} _\text{new} $ by Eq. (\ref{eq37});\\
				\STATE Compute the current residual $\mathbf{r} (\mathbf{p}_{\text{new}} )$ by Eq. (\ref{eq38});\\
				\STATE Calculate $ \|\mathbf{r} (\mathbf{p}_{\text{new}} )\|$ and $ \|\mathbf{r} (\mathbf{p}_j )\|$;\\
				\IF{$ \|\mathbf{r} (\mathbf{p}_{\text{new}} )\| <  \|\mathbf{r} (\mathbf{p}_j )\|$}
				\STATE $\lambda _{j+1} \gets \lambda _j/\gamma$; \quad $\mathbf{p} _{j+1}\gets \mathbf{p} _\text{new}$ \quad $\mathbf{r} (\mathbf{p}_{j+1} ) \gets \mathbf{r} (\mathbf{p}_{\text{new}} )$;\\
				\ELSE
				\STATE $\lambda _{j+1} \gets \gamma \, \lambda _j$; \quad $\mathbf{p} _{j+1}\gets \mathbf{p} _j$ \quad $\mathbf{r} (\mathbf{p}_{j+1} ) \gets \mathbf{r} (\mathbf{p}_{j} )$;\\
				\ENDIF
				\ENDFOR
			}
			
			\STATE {\textbf{Step 3: Parameter estimation}\\
				\STATE Split $\mathbf{p} _{J}$ into $\hat{a} _k$ and $\hat{b} _k$
				\STATE $\hat{A}_k (t_l) \gets |\hat{a} _k|,\quad \hat{\varphi}_k(t_l) \gets \angle \hat{a} _k , \quad \hat{\eta}_k  \gets \frac{\Im \{\hat{b} _k\}}{2\pi}$;\\
				\STATE $\hat{f}^{\text{final}}_k(t_l) \gets \hat{f}_k(t_l) + \hat{\eta}_k $
			}
			\STATE {\textbf{Output:} $\hat{A}_k (t_l)$, $\hat{\varphi}_k(t_l)$, $\hat{f}^{\text{final}}_k(t_l)$}
		\end{algorithmic}
		\label{high-order-SM}
	\end{algorithm}

	\vspace{-0.3cm}
	
	\subsection{Signal Reconstruction}
	Second, similar to QHM methods, by correcting the frequencies and adopting interpolations for framewise amplitude and frequency, their instantaneous versions $\hat{A}_k(t)$ and $\hat{f}^{\text{final}}_k(t)$ can be smoothly depicted to obtain the instantaneous phase $\hat{\varphi}_k(t)$ in a special way \cite{pantazis2010adaptive}, \cite{bpqhm}:
	\begin{align}
	\scalebox{0.94}{${{\hat{\varphi }}_{k}}(t)={{\hat{\varphi }}_{k}}({{t}_{l}})+\int_{{{t}_{l}}}^{t_l+t}{2\pi \hat{f}^{\text{final}}_k(u)}+z\sin \left[ \frac{\pi (u-{{t}_{l-1}})}{{{t}_{l}}-{{t}_{l-1}}} \right]\text{d}u,$}
	\label{eq41_2}
	\end{align}
	where ${{\hat{\varphi }}_{k}}({{t}_{l}}) $ equals the $ \hat{\phi}_k$ of $l$-th frame from Eq. (\ref{eq21}) and $z$ is computed by
	\begin{align*}
	z=\frac{\pi [{{{\hat{\varphi }}}_{k}}({{t}_{l+1}})+2\pi \mathcal{M}-{{{\tilde{\varphi }}}_{k}}({{t}_{l+1}})]}{2({{t}_{l+1}}-{{t}_{l}})}.
	\end{align*}
	$\tilde{\varphi }_{k}({{t}_{l+1}})$ is obtained by Eq. (\ref{eq8}) and $\mathcal{M}$ is the closest integer to ${\left| {{{\hat{\varphi }}}_{k}}(t_{l+1})-{{{\tilde{\varphi }}}_{k}}(t_{l+1}) \right|}/{2\pi }]$. Then, arbitrary component can be reconstructed by
	\begin{align} 
		\hat{x}_k(t)= \hat{A}_k(t) e^{i \hat{\varphi}_k(t)}.
		\label{eq41}
	\end{align}

	\vspace{-0.3cm}
	
	\section{Numerical Performance Evaluation}
	To evaluate the performance of the proposed high-order SM, a series of experiments were conducted to compare 4th-SM with QHM and eaQHM in terms of reconstruction accuracy and estimation accuracy for amplitudes and frequencies. Moreover, to examine the effect of the model order on performance, SMs of different orders (2nd, 4th, and 6th) were employed to examine the speech. The experiments were based on the analysis of the following simulated multicomponent signal:
	\vspace{-0.1cm}
	\begin{alignat}{2}\label{eq42}
	x(t)&=\sum_{k=1}^{6} A_k(t)e^{i2\pi\int_0^t f_k(u)\,du}, &\quad& t \in [0,1], \quad \text{where} \\
	A_1 &= 0.18 + 0.1 \sin(2\pi t), & f_1 &= 200, \notag\\
	A_2 &= 0.2 + 0.1 \sin(2\pi t), & f_2 &= 400, \notag\\
	A_3 &= 0.15 - 0.1 \cos(2\pi t), & f_3 &= 2000 + 300 t, \notag\\
	A_4 &= 0.3 - 0.05 \cos(2\pi t), & f_4 &= 2200 + 400 t, \notag\\
	A_5 &= 0.08 + 0.1 t, & f_5 &= 4000 + 50 \cos(10\pi t), \notag\\
	A_6 &= 0.25 - 0.15 t, & f_6 &= 4250 + 50 \cos(10\pi t), \notag
	\end{alignat}
	which contains diverse and complex amplitude and frequency modulations. Two cases were considered in the experiments: (i) noisy frequency and (ii) noisy frequency and waveform.
	
	To quantify the performance of all methods, we employ the root-mean-squared errors
	(RMSE) to measure the estimation accuracies of frequency and amplitude, as 
	\begin{align} \label{eq43}
	\text{RMSE}_A &= \sqrt{\frac{1}{L}\sum\limits_{l=1}^{L}{\left[ A{{\left( t_l \right)}}-\hat{A}{{\left( t_l \right)}} \right]^{2}}}, \\
	\text{RMSE}_f &= \sqrt{\frac{1}{L}\sum\limits_{l=1}^{L}{\left[ \ln f{{\left( t_l \right)}}-\ln \hat{f}{{\left( t_l \right)}} \right]^{2}}},
	\label{eq44}
	\end{align}
	and the signal-to-reconstruction-error ratio (SRER) to measure the reconstruction quality, which is formulated as
	\vspace{-0.2cm}
	\begin{align}
	\text{SRER} = 20\log_{10}\frac{\text{std}(x)}{\text{std}(x-\hat{x})}, 
	\label{eq45}
	\end{align}
	where $\text{std}(\cdot)$ is the standard deviation and a larger SRER denotes a result closer to the reference.

	\vspace{-0.1cm}
	\subsection{Case 1: Noisy Frequency Scenario}
\begin{figure}[!t]\centering
		\includegraphics[width=8.89cm]{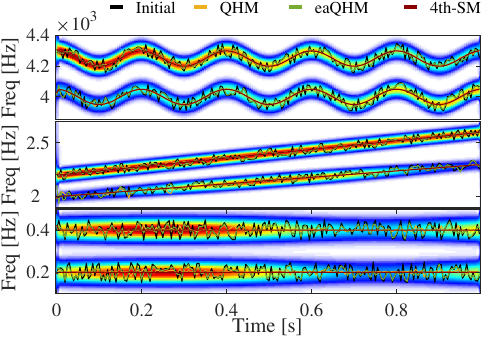}
		\vspace{-0.7cm}
		\caption{Frequency trajectories estimated by all methods.}
		\label{FIG8}
		\vspace{-0.5cm}
	\end{figure}

	To assess frequency correction and parameter estimation accuracy, the frequencies of all components were randomly perturbed by $\eta_k \in [-\mu, \mu]$ ($\mu = 50$) and used as initial frequencies for QHM, eaQHM, and 4th-SM. The spectrogram with the estimated trajectories is shown in Fig. \ref{FIG8}, and Table \ref{table1} reports $\text{RMSE}_A$ and $\text{RMSE}_f$ for each component.
	
	Jointly analyzing Fig. \ref{FIG8} and Table \ref{table1}, several observations arise. First, QHM partially alleviates frequency mismatch, yet its performance deteriorates under large initial mismatches. QHM also yields the highest $\mathrm{RMSE}_A$ for all components, reflecting its limited accuracy in amplitude estimation caused by the low model expressiveness. In contrast, eaQHM provides improved amplitude estimation owing to the introduction of an amplitude amplifier, yet fails to effectively correct the frequency mismatch and may even aggravate them in some cases. On the other hand, 4th-SM consistently outperforms both QHM and eaQHM in terms of the frequency trajectories, $\mathrm{RMSE}_A$, and $\mathrm{RMSE}_f$. This improvement benefits from the high-order formulation of the 4th-SM, which enhances the representational capacity for both the amplitude and the frequency mismatch terms, thereby providing a more faithful parameter estimation.

	\begin{table}[!t]
		\caption{Amplitude and frequency RMSE scores for each component of all methods: $\text{RMSE}_A \downarrow$ and $\text{RMSE}_f \downarrow \, (\times 10^{-3})$.}
		\centering
		\setlength\tabcolsep{1pt} 
		\label{table1}
		\begin{tabular}{c|cccccc}
			\toprule[1pt]
			$\text{RMSE}_A / \text{RMSE}_f $ & $1$st & $2$nd & $3$rd  & $4$th & $5$th & $6$th \\\midrule[0.5pt]
			QHM     & 19.3/28.9 & 19.4/13.7 & 13.4/4.5  & 26.7/1.6 & 11.6/1.1 & 16.7/0.6 \\\midrule[0.5pt]
			eaQHM   & 15.9/99.3 & 15.7/45.8 & 12.5/10.1  & 16.2/7.8 & 7.4/4.7 & 13.2/4.2 \\\midrule[0.5pt]
			4th-SM  & 5.9/3.2 & 5.9/3.9 & 0.5/0.5  & 10.5/0.3 & 1.9/0.4 & 8.6/0.2 \\
			\bottomrule[1pt] 
		\end{tabular}
	\vspace{-0.3cm}
	\end{table}
	
\begin{figure}[!t]\centering
		\includegraphics[width=8.89cm]{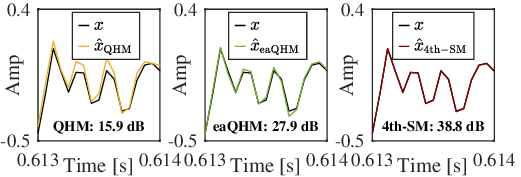}
		\vspace{-0.7cm}
		\caption{The signal reconstructed by all methods.}
		\label{FIG9}
		\vspace{-0.5cm}
	\end{figure}

	Next, the reconstruction ability is examined. Figure \ref{FIG9} shows local waveform segments around $t = 0.613$s, comparing the ground truth with the synthesized waveforms from different methods. Obviously, QHM and eaQHM exhibit noticeable deviations, whereas 4th-SM produces a waveform closer to the ground truth. The corresponding SRER values in Fig. \ref{FIG9} also confirm this trend. Compared with QHM, eaQHM benefits from the incorporation of the amplitude amplifier and time-varying frequency across iterations, leading to more accurate reconstructions. The 4th-SM achieves the highest SRER value (38.8 dB), which can be attributed to its precise parameter estimation, yielding highly accurate waveform reconstructions.

\begin{figure}[!t]\centering
		\includegraphics[width=8.89cm]{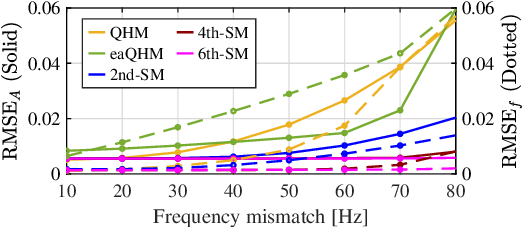}
		\vspace{-0.7cm}
		\caption{$\text{RMSE}_A$ and $\text{RMSE}_f$ scores of all methods under different frequency mismatch conditions.}
		\label{FIG10}
		\vspace{-0.4cm}
	\end{figure}

\begin{figure}[!t]\centering
		\includegraphics[width=8.89cm]{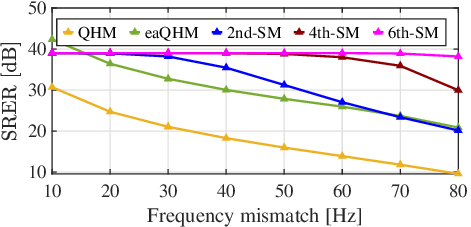}
		\vspace{-0.7cm}
		\caption{SRER scores of all methods under different frequency mismatch conditions.}
		\label{FIG11}
		\vspace{-0.5cm}
	\end{figure}

	To investigate the robustness against varying frequency mismatches, the frequencies were corrupted with random mismatches ($\mu = 10,\dots,80$) for 20 Monte-Carlo simulations. Moreover, 2nd- and 6th-SMs were additionally evaluated to highlight the performance trend as the model order increases. Figures \ref{FIG10} and \ref{FIG11} show the average curves of $\mathrm{RMSE}_A$, $\mathrm{RMSE}_f$, and SRER, respectively. As the initial frequency mismatch increases, QHM and eaQHM exhibit lower and progressively degraded accuracy in amplitude and frequency estimation as well as reconstruction, whereas the 2nd- and 4th-SMs achieve better yet slightly declining performance under larger mismatches. Notably, 4th-SM outperforms 2nd-SM. The 6th-SM maintains stable performances across all mismatch levels, confirming that increasing the model order consistently enhances estimation accuracy and reconstruction fidelity.

	\vspace{-0.3cm}
	\subsection{Case 2: Noisy Frequency and Noisy Waveform Scenario}
	
\begin{figure}[!t]\centering
		\includegraphics[width=8.89cm]{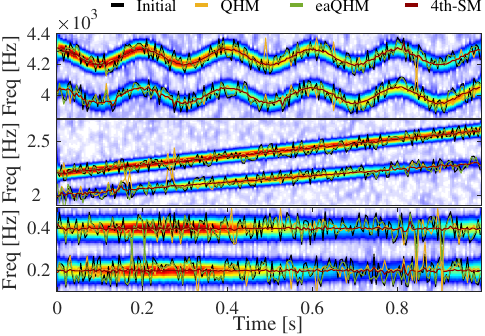}
		\vspace{-0.7cm}
		\caption{Frequency trajectories estimated by all methods under noisy cases.}
		\label{FIG12}
		\vspace{-0.2cm}
	\end{figure}

\begin{figure}[!t]\centering
		\includegraphics[width=8.89cm]{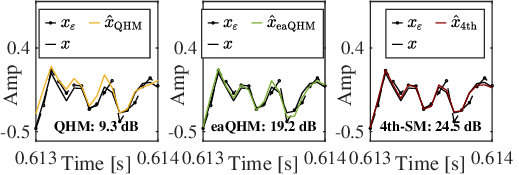}
		\vspace{-0.7cm}
		\caption{The signal reconstructed by all methods under noise case.}
		\label{FIG13}
		\vspace{-0.4cm}
	\end{figure}

	In this part, we investigate the robustness of the methods to waveform noise. Specifically, all methods are initialized with frequencies corrupted by the mismatch ($\mu = 80$) and applied to the signal corrupted with additive Gaussian white noise at signal-to-noise ratio (SNR) equals 15 dB, where
	\begin{align}
	\text{SNR} = 20\log_{10}\frac{\text{std}(x)}{\text{std}(\varepsilon)}.
	\label{eq46}
	\end{align}
	The frequency trajectories, parameter estimation accuracy, and reconstruction performance of QHM, eaQHM, and 4th-SM are presented in Fig.~\ref{FIG12}, Table~\ref{table2}, and Fig.~\ref{FIG13}, respectively.

	Clearly, the presence of noise affects each method to varying degrees. Fig. \ref{FIG12} shows that the frequency trajectories of QHM and eaQHM often deviate from the ground truth and sometimes erroneously swap between neighboring components. Contrarily, 4th-SM exhibits much smaller deviations, showing the best alignment with true trajectories. Table \ref{table2} further highlights that QHM and eaQHM incur substantial amplitude estimation errors, due to their low-order nature. In particular, the amplitude-sensitive design of eaQHM makes it more prone to noise, resulting in even larger $\text{RMSE}_A$ than QHM in the first two components. In contrast, 4th-SM benefits from its higher-order structure, offering enhanced model stability and stronger noise robustness. Thus, 4th-SM achieves much lower $\text{RMSE}_A$ and $\text{RMSE}_f$ across all components.
	

	\begin{table}[!t]
		\caption{Amplitude and frequency RMSE scores for each component of all methods under noise case: $\text{RMSE}_A \downarrow$ and $\text{RMSE}_f \downarrow \, (\times 10^{-3})$.}
		\centering
		\setlength\tabcolsep{0.5pt} 
		\label{table2}
		\begin{tabular}{c|cccccc}
			\toprule[1pt]
			$\text{RMSE}_A / \text{RMSE}_f $ & $1$st & $2$nd & $3$rd  & $4$th & $5$th & $6$th \\\midrule[0.5pt]
			QHM     & 66.3/150.7 & 65.2/137.5 & 35.5/15.1  & 72.4/9.7 & 33.2/5.5 & 42.9/5.1 \\\midrule[0.5pt]
			eaQHM   & 121.8/193.7 & 102.4/94.5 & 24.1/17.5 & 25.0/12.9 & 18.3/8.0 & 19.5/7.3 \\\midrule[0.5pt]
			4th-SM  & 10.5/24.6 & 10.6/11.9 & 9.5/3.7 & 13.1/1.6 & 11.6/2.0 & 12.3/1.3 \\
			\bottomrule[1pt] 
		\end{tabular}
	\vspace{-0.3cm}
	\end{table}

	Similar to Fig. \ref{FIG9}, Fig. \ref{FIG13} examines the reconstruction performance using local waveform segments around $t = 0.613$s. All methods achieve partial denoising due to their sinusoidal nature. However, the waveform reconstructed by QHM exhibits the largest deviations from the ground truth, whereas eaQHM achieves moderate improvements, though noticeable errors remain. The reconstructed signal of 4th-SM closely matches the ground truth, as also reflected by the SRER values (24.5 dB) in Fig. \ref{FIG13}. Note that, although eaQHM achieves a higher SRER value than QHM, it exhibits substantially higher $\text{RMSE}_A$ and $\text{RMSE}_f$ due to iterative adjustments of frequencies and amplitudes across frames for approximating true waveform, which can compromise parameter accuracy. This highlights the limitation of eaQHM: it prioritizes waveform reconstruction over precise parameter estimation. In contrast, 4th-SM achieves both accurate parameter estimation and faithful signal reconstruction.

\begin{figure}[!t]\centering
		\includegraphics[width=8.89cm]{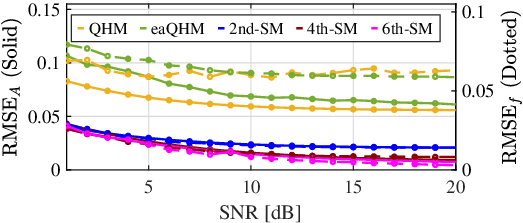}
		\vspace{-0.7cm}
		\caption{$\text{RMSE}_A$ and $\text{RMSE}_f$ scores of all methods under different noise levels.}
		\label{FIG14}
		\vspace{-0.4cm}
	\end{figure}

\begin{figure}[!t]\centering
		\includegraphics[width=8.89cm]{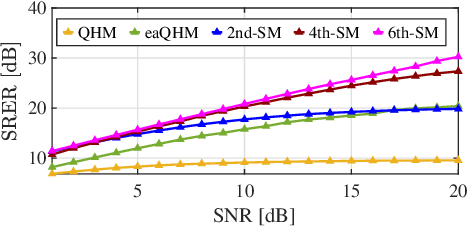}
		\vspace{-0.7cm}
		\caption{SRER scores of all methods under different noise levels.}
		\label{FIG15}
		\vspace{-0.5cm}
	\end{figure}

	To further investigate the performance under varying noise levels, the $\eta_k $ is randomly initialized at $\mu = 80$, and additive Gaussian noise with SNRs from 1 dB to 20 dB was applied in 20 Monte-Carlo simulations. All methods, along with the additionally included 2nd- and 6th-SMs, are then applied to extract the parameters, where the average curves of $\mathrm{RMSE}_A$, $\mathrm{RMSE}_f$, and SRER are presented in Figs. \ref{FIG14} and \ref{FIG15}. As the noise level increases, the estimation and reconstruction errors of all methods increase correspondingly. Notably, the high-order SM consistently exhibits smaller errors than the QHM methods. Among them, the 6th-SM demonstrates the highest robustness to noise, achieving the best SRER and the lowest estimation errors across all noise levels.
	
	\vspace{-0.2cm}

	\section{Application in Speech Modeling}
	
	Speech modeling plays a fundamental role in modern speech processing, serving as the basis for representing acoustic features and reconstructing signals, enabling applications such as synthesis, modification, and voice conversion. The accuracy of parameter estimation, particularly frequency, is critical since deviations directly affect the quality of subsequent processing. Conventional frequency estimation typically relies on $f_0$ detectors, which inevitably introduce estimation errors. QHM methods address this issue by jointly estimating complex amplitudes and correcting frequencies on a frame-by-frame basis, making them highly suitable for speech modeling. In this section, we compare high-order SM with the QHM methods in terms of frequency and amplitude estimation accuracy.

	During the experiments, speech utterances randomly selected from LJspeech \cite{ljspeech17} (22.05 kHz) were analyzed with the $f_0$ initialized by YAAPT \cite{kasi2002yet}. Since the ground truth of individual amplitudes and frequencies is unavailable, RMSE cannot be directly applied to evaluate the estimation accuracy. Following the idea in \cite{bpqhm}, the spectrogram of a speech signal can be approximated by the extracted parameters as
	\begin{align}
	\hat{S}_x(t,\omega) \approx \sum_{k=-K}^{K}A_k(t)G(\omega_k - \omega)e^{i\omega_k t},
	\label{eq47}
	\end{align}
	\noindent where $\omega_k = 2\pi f_k(t)$ and $G(\cdot)$ denotes the Fourier transform of the Gaussian window. Therefore, the distance between Eq. (\ref{eq47}) and the spectrogram of the original signal can serve as a metric for parameter estimation accuracy. This distance, denoted as $L_{\text{spec}}$, is defined as
	\begin{align}
	L_{\text{spec}} = \frac{1}{L}\sum_{l = 1}^{L}\sqrt{\frac{1}{N_f} \sum_{n_f = 1}^{N_f}(\scalebox{0.97}{$|\hat{S}_x(t_l,n_f)| - |S_x(t_l,n_f)|$})^2},
	\label{eq48}
	\end{align}
	where $S_x$ is the reference spectrogram and $N_f$ denotes the half size of the Fourier transform. A lower $L_{\text{spec}}$ indicates a more accurate estimation of both amplitude and frequency.

\begin{figure}[!t]\centering
		\includegraphics[width=8.89cm]{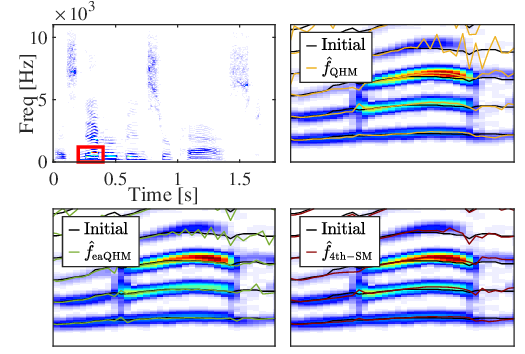}
		\vspace{-0.7cm}
		\caption{The spectrogram and the frequency trajectories by all methods.}
		\label{FIG16}
		\vspace{-0.4cm}
	\end{figure}

	To balance accuracy and computational cost, the 4th-SM was adopted in the following experiments. Figure \ref{FIG16} illustrates the spectrogram of a speech sample together with the frequency trajectories estimated by all methods. QHM provides reasonably accurate estimates, yet serrated trajectories appear around the 4th--6th harmonics. eaQHM alleviates such serrations, however, it fails to reliably correct the frequencies of the lower harmonics (1st--3rd). In contrast, 4th-SM achieves consistently accurate frequency estimation across all harmonics, preliminarily showing the superior frequency estimation.  
	
\begin{figure}[!t]\centering
		\includegraphics[width=8.89cm]{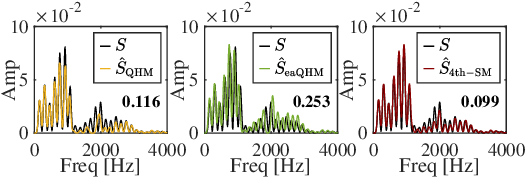}
		\vspace{-0.7cm}
		\caption{The spectrograms of all methods generated by Eq. (\ref{eq47}).}
		\label{FIG17}
		\vspace{-0.4cm}
	\end{figure}

	Additionally, Fig. \ref{FIG17} compares the spectrograms of all methods generated by Eq. (\ref{eq47}) against the ground truth, together with the corresponding $L_{\text{spec}}$ around $t = 1.16$s. QHM yields relatively accurate frequencies but suffers from large amplitude errors. Conversely, eaQHM estimates amplitudes more precisely but produces noticeable frequency deviations, leading to a large $L_{\text{spec}}$. 4th-SM achieves both accurate amplitude and frequency estimates, attaining the smallest $L_{\text{spec}}$.  
	
\begin{figure}[!t]\centering
		\includegraphics[width=8.89cm]{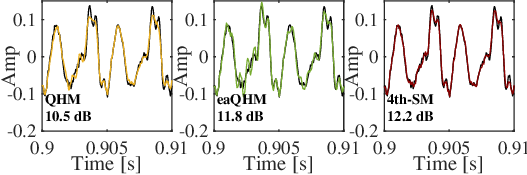}
		\vspace{-0.7cm}
		\caption{Reconstructed waveforms and corresponding SRER of all methods.}
		\label{FIG18}
		\vspace{-0.5cm}
	\end{figure}

	Figure \ref{FIG18} further shows the reconstructed waveforms and SRER. A similar trend can be observed: QHM captures frequencies but poorly models amplitudes, while eaQHM emphasizes amplitude accuracy at the cost of frequency estimation, making it sensitive to waveform perturbations. Due to the accurate estimation in both amplitude and frequency, 4th-SM attains the most reliable reconstruction, yielding the highest SRER of 12.2~dB. These results demonstrate the robustness of 4th-SM in speech modeling.

	For a more comprehensive comparison, we conducted experiments on random 32 utterances. The averaged results are listed in Table \ref{table3}, including $L_\text{spec}$, SRER, and real-time factor (RTF), which quantifies the time consumption for 1 sec signal using an Intel i7 8700K CPU. Since QHM methods\footnote{https://github.com/Antibas/eaQHM-analysis-and-synthesis-in-Python/tree/main} are established without parallel computation, for fair comparison, all methods were implemented using a single CPU core and processed frame by frame. The results demonstrate that 4th-SM consistently achieves the best performance in both parameter estimation accuracy and reconstruction fidelity, thereby confirming the superiority of the higher-order SM. Regarding the efficiency, due to the iterations in LM and eaQHM, 4th-SM and eaQHM achieve higher RTFs than QHM. Notably, in 4th-SM, the number of LM iterations is fixed and typically small (usually 8), whereas in eaQHM the iteration number is not fixed. Consequently, when the initial frequency mismatch is large, more iterations are required, leading to higher computational cost, as reflected by the RTF value approaching 26 in the table.

	\begin{table}[!t]
		\caption{Average $L_\text{spec}$, SRER, and RTF scores of all methods.}
		\centering
		\setlength\tabcolsep{4pt} 
		\label{table3}
		\begin{tabular}{c|ccc}
			\toprule[1pt]
				Methods			 & QHM & eaQHM  & $4$th-SM  \\\midrule[0.5pt]
			$L_\text{spec}$ $\downarrow$ $\times 10^{-2}$  & 10.96 & 26.31 & 9.38  \\\midrule[0.5pt]
			SRER [dB] $\uparrow$       & 8.8 & 11.5 & 12.7   \\\midrule[0.5pt]
			RTF $\downarrow$  & 4.39 & 25.49 & 9.49   \\
			\bottomrule[1pt] 
		\end{tabular}
	\vspace{-0.5cm}
	\end{table}

	\vspace{-0.1cm}
	\section{Conclusion}
	In this paper, the principle and limitations of QHM methods are reviewed, showing their inaccuracy in parameter extraction and reconstruction, especially under large frequency mismatches. To improve the accuracy, we first re-derive the QHM methods to dig out the main reason for the inaccuracy. Second, we increase the order of the model and propose a high-order sinusoidal model (SM) to improve the expressive ability of the model to the frequency mismatches and amplitudes. Eventually, we employ the Levenberg-Marquardt algorithm to adaptively obtain the parameters with fewer iterations, achieving markedly faster and more accurate estimation. The results of simulation study and experiments in speech modeling indicated that high-order SM can more accurately estimate the amplitudes and frequencies of all components in signals with limited time consumption (faster than eaQHM), accordingly improving the reconstruction ability and the robustness against the waveform noise and frequency perturbance. Thus, high-order SM shows the promise for applications in the fine-grained analysis of signals.

	\appendices
	\renewcommand{\thesection}{\Alph{section}}
	\section{The Proof of Theorem 1}
	\label{app:theorem1}
	
	\emph{Proof:} Let $E(a,b) = \left\| x^{(N)}(t; a, b)- x(t) \right\|^2$ and denoting $x(t)$ and $y^{(N)}(t; b)$ as $\mathbf{x}$ and $\mathbf{y}_N$, respectively, Eq. (\ref{eq23}) reads
	\begin{align}
		(\hat{a}, \hat{b}) = \arg\min_{a, b} \frac{1}{2}E(a,b) = \arg\min_{a, b} \frac{1}{2}\left\| a \mathbf{y}_N- \mathbf{x} \right\|^2.
		\label{Aeq1}
	\end{align}
	The coefficient $a$ corresponds to the orthogonal projection of $\mathbf{x}$ onto the subspace spanned by the basis function $\mathbf{y}_N$. Thus, according to ${\partial E(a,b)}/{\partial a^*}=0 $, we have 
	\begin{align}
	  a = \frac{\langle \mathbf{x},\mathbf{y}_N \rangle}{\langle \mathbf{y}_N,\mathbf{y}_N\rangle}= \frac{\langle x(t),y^{(N)}(t; b) \rangle}{\left\| y^{(N)}(t; b) \right\|^2},
	\label{Aeq2}
	\end{align}
	where $\langle \mathbf{x}, \mathbf{y} \rangle = \mathbf{y}^\dagger \mathbf{x}$ is the complex inner product and $a$ is a complex-valued function of $b$. The cost function $E(a,b)$ is nonlinear in $b$ due to the exponential-like polynomial structure of $y^{(N)}(t; b)$. Substituting Eq. (\ref{Aeq2}) into $E(a,b)$ yields:
	\begin{align}
	E(a,b) &= \left\|  a \mathbf{y}_N- \mathbf{x} \right\|^2 = \langle  a \mathbf{y}_N- \mathbf{x}, a \mathbf{y}_N- \mathbf{x}\rangle \notag\\
	& = aa^*\langle \mathbf{y}_N,\mathbf{y}_N \rangle - a\langle \mathbf{y}_N,\mathbf{x} \rangle -a^*\langle \mathbf{x} ,\mathbf{y}_N\rangle + \langle \mathbf{x},\mathbf{x} \rangle \notag \\
	&= \langle \mathbf{x},\mathbf{x} \rangle - \frac{ \left| \langle \mathbf{x},\mathbf{y}_N \rangle  \right|^2 }{\langle \mathbf{y}_N,\mathbf{y}_N \rangle} = \left\|  \mathbf{x} \right\|^2 - \frac{ \left| \langle \mathbf{x},\mathbf{y}_N \rangle  \right|^2}{\left\|  \mathbf{y}_N \right\|^2}.
	\label{Aeq3}
	\end{align}
	Since there exists a constant $C \in \mathbb{R^+}$ such that $\left\| \mathbf{x} \right\|^2 > C$, the solution of $b$ can be equivalently rewritten as follows:
	\begin{align}
	\hat{b} = \arg\min_{b}\frac{1}{2} E(b) = \arg\max_{b}\frac{ \left| \langle \mathbf{x},\mathbf{y}_N \rangle  \right|^2}{\left\|  \mathbf{y}_N \right\|^2}.
	\label{Aeq4}
	\end{align}
	According to the result of $\hat{a}$ and $\hat{b}$, the amplitude and frequency mismatch can be estimated by
	\begin{align}
	\hat{A} &= |\hat{a}|=\frac{\left| \langle x(t), y^{(N)}(t; b) \rangle\right|}{\left\|y^{(N)}(t; b)\right\|^2}, \\
	\hat{\eta} &= 	\frac{\Im \left\{ \hat{b} \right\}}{2\pi} = \Im \left\{\arg\max_{b} \scalebox{1.225}{$ \frac{|\langle x(t), y^{(N)}(t; b) \rangle|^2}{\left\|y^{(N)}(t; b)\right\|^2} $} \right\} /2\pi.
	\label{Aeq5}
	\end{align}
	Then, we complete the proof of Theorem 1.
	
	\section{The Proof of Proposition 1}
	\label{app:proposition1}

	Here, we analyze the estimation accuracy of the model parameters. Similar to refs.\cite{katkovnik1998instantaneous},\cite{sejdic2008quantitative},\cite{wang2013matching},\cite{chen2023instantaneous}, which estimate the instantaneous frequency, our analysis also relies on the optimality conditions ${\partial E(a,b)}/{\partial a^*}=0$ and ${\partial E(a,b)}/{\partial b^*}=0$ to estimate $a$ and $b$. Since $E(a,b)$ is too complex, we linearize it around the point ($a^\star=A e^{i\phi}, b^\star = i2\pi\eta$) with respect to small estimation errors $\varepsilon_a$ and $\varepsilon_b$. The linearization reads
	\begin{align}\label{Aeq6_1}
	\frac{\partial E(a,b)}{\partial a^*} &= \left. \frac{\partial E(a,b)}{\partial a^*}\right|_{\substack{a=a^\star \\  b = b^\star}} + \left. \frac{\partial^2 E(a,b)}{\partial a \partial a^*}\right|_{\substack{ b = b^\star}}\varepsilon_a = 0, \\
	\frac{\partial E(a,b)}{\partial b^*} &\approx \left. \frac{\partial E(a,b)}{\partial b^*}\right|_{\substack{a=a^\star \\  b = b^\star}}   +  \left.	\frac{\partial^2 E(a,b)}{\partial b \partial b^*}\right|_{\substack{a=a^\star \\  b = b^\star}}\varepsilon_b = 0.
	\label{Aeq6_2}
	\end{align}
	Now, we separately analyze the estimation accuracy of $a$ and $b$. First, we start from on the amplitude term.  According to Eq. (\ref{Aeq6_1}), we have 
	\begin{align}
	&\varepsilon_a = \frac{\langle \mathbf{x} ,\mathbf{y}_N^\star\rangle - A e^{i\phi}\langle \mathbf{y}_N^\star,\mathbf{y}_N^\star \rangle }{\langle \mathbf{y}_N^\star,\mathbf{y}_N^\star \rangle} = \frac{\langle \mathbf{x} ,\mathbf{y}_N^\star\rangle}{\langle \mathbf{y}_N^\star,\mathbf{y}_N^\star \rangle} - A e^{i\phi} \notag \\
	&=  \frac{A e^{i\phi} \langle \mathbf{y}_N^\star + \mathbf{R}_N^\star ,\mathbf{y}_N^\star\rangle}{\langle \mathbf{y}_N^\star,\mathbf{y}_N^\star \rangle} - A e^{i\phi} = \frac{A e^{i\phi} \langle \mathbf{R}_N^\star ,\mathbf{y}_N^\star\rangle}{\|\mathbf{y}_N^\star\|^2}
	\label{Aeq7}
	\end{align}
	where $\mathbf{y}_N^\star = y^{(N)}(t; i2\pi\eta)$.  $ \mathbf{x}$ is expanded into a Taylor series $\mathbf{y}_N^\star$ and a Lagrange remainder $ \mathbf{R}_N^\star$, i.e., $ \mathbf{x} = Ae^{i\phi}[\mathbf{y}_N^\star + \mathbf{R}_N^\star ]$. According to Cauchy-Schwarz inequality, we have 
	\begin{align}
	\left|\varepsilon_a\right| = A\frac{ | \langle \mathbf{R}_N^\star ,\mathbf{y}_N^\star\rangle |}{\|\mathbf{y}_N^\star\|^2} \leq A\frac{ \|\mathbf{R}_N^\star \| \| \mathbf{y}_N^\star \|}{\|\mathbf{y}_N^\star\|^2} = A\frac{ \|\mathbf{R}_N^\star \|}{\|\mathbf{y}_N^\star\|}.
	\label{Aeq8}
	\end{align}
	Denoting the amplitude estimation error as $\varepsilon_A$, we have $\left|\varepsilon_A\right| = \left|\varepsilon_a\right|$. Since $ \mathbf{R}_N^\star = \frac{(i2\pi\eta t)^{N+1}e^{i \xi}}{(N+1)!}  e^{i2\pi \hat{f}t} w(t), \xi \in [0,2\pi\eta t]$ and $\exists C_2 \in \mathbb{R^+},\ \| \mathbf{y}_N^\star \| > C_2>0$, Eq. (\ref{Aeq8}) can obtain:
	\begin{align}\small
	\scalebox{1.145}{$\left|\varepsilon_A\right| \leq \frac{A \|\mathbf{R}_N^\star \|}{\|\mathbf{y}_N^\star\|} \leq \frac{A|2\pi\eta|^{N+1}}{C_2(N+1)!}\sqrt{\int_{-T}^{T}t^{2(N+1)}w^2(t)dt} $}.
	\label{Aeq9}
	\end{align}
	This result indicates that the estimation error decreases with increasing the order $N$, and approaches zero in the limit.

\begin{figure}[!t]\centering
		\includegraphics[width=8.89cm]{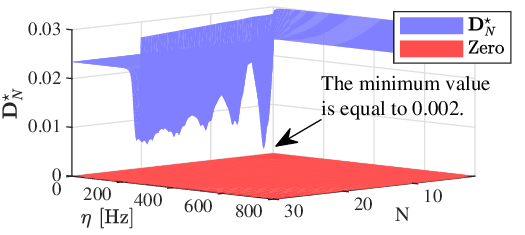}
		\vspace{-0.7cm}
		\caption{$\mathbf{D}_N^\star$ in Eq. (\ref{Aeq11}) as a function of the initial frequency mismatch $\eta$ and the model order $N$ ($T=10$ms).}
		\label{FIG19}
		\vspace{-0.7cm}
	\end{figure}

	Likewise, from Eq. (\ref{Aeq6_2}), we can analyze the estimation accuracy of $b$ (denoted as $\varepsilon_b $), as
	\begin{align}
	\varepsilon_b &= - \frac{\langle A e^{i\phi}\mathbf{y}_N^{'\star},A e^{i\phi}\mathbf{y}_N^\star -  \mathbf{x} \rangle}{\langle  A e^{i\phi}\mathbf{y}_N^{''\star},A e^{i\phi}\mathbf{y}_N^\star  -  \mathbf{x}\rangle + \|Ae^{i\phi}\mathbf{y}_N^{'\star}\|^2} \notag \\
	&= \frac{ \langle \mathbf{t} \circ \mathbf{y}_{N-1}^\star,\mathbf{R}_N^\star  \rangle}{\|\mathbf{t} \circ \mathbf{y}_{N-1}^\star\|^2 -\langle \mathbf{t}^{\circ 2} \circ \mathbf{y}_{N-2}^\star,\mathbf{R}_N^\star \rangle}
	\label{Aeq10}
	\end{align}
	where $\mathbf{y}_N^{'\star}$ and $\mathbf{y}_N^{''\star}$ denote the first- and second-order partial derivatives of $\mathbf{y}_N$ with respect to $b$, evaluated at $b = i2\pi\eta$. $\mathbf{y}_N^{'\star} = \mathbf{t} \circ\mathbf{y}_{N-1}^\star $ ($\circ$ is the Hadamard product) is considered. Since the estimation error of $\eta$, denoted as $\varepsilon_\eta$, satisfies $\left|\varepsilon_\eta\right| = \frac{\left|\Im \{\varepsilon_b\}\right|}{2\pi} \leq \frac{\left|\varepsilon_b\right|}{2\pi}$, Cauchy-Schwarz inequality yields
	\begin{align}
	2\pi\left|\varepsilon_\eta\right| \leq  \scalebox{1.2}{$\frac{ \| \mathbf{t} \circ \mathbf{y}_{N-1}^\star\| \|\mathbf{R}_N^\star  \|}{\|\mathbf{t} \circ \mathbf{y}_{N-1}^\star\|^2 -\| \mathbf{t}^{\circ 2} \circ \mathbf{y}_{N-2}^\star\| \|\mathbf{R}_N^\star \|} = \frac{ \|\mathbf{R}_N^\star  \|}{\mathbf{D}_N^\star }$},
	\label{Aeq11}
	\end{align}
	where $\mathbf{D}_N^\star = \|\mathbf{t} \circ \mathbf{y}_{N-1}^\star\| - \| \mathbf{t}^{\circ 2} \circ \mathbf{y}_{N-2}^\star\| \|\mathbf{R}_N^\star \| / \|\mathbf{t} \circ \mathbf{y}_{N-1}^\star\|> 0$ is required to guarantee the stability of the estimation. Although a closed-form analytical proof of $\mathbf{D}_N^\star > 0$ is intractable, empirical results show that it holds for a wide range of $\eta$ and $N$ when using symmetrically tapered windows. An example with Blackman window is illustrated in Fig. \ref{FIG19}, where the minimum of $\mathbf{D}_N^\star$ stays strictly positive.	
	Since $\exists C_3 \in \mathbb{R^+}, \mathbf{D}_N^\star > C_3$, Eq. (\ref{Aeq11}) yields:
	\begin{align}
	\left|\varepsilon_\eta\right| \leq \frac{ \|\mathbf{R}_N^\star \|}{\mathbf{D}_N^\star} \leq \frac{|2\pi|^{N}|\eta|^{N+1}}{C_3(N+1)!}\sqrt{\int_{-T}^{T}t^{2(N+1)}w^2(t)dt}.
	\label{Aeq12}
	\end{align}
	Then we complete the proof of Proposition 1.
	
	\bibliographystyle{IEEEtran}
	\bibliography{IEEEtranTSP}
\end{document}